\documentclass[twocolumn,showpacs,preprintnumbers,amsmath,amssymb]{revtex4}
\usepackage{graphicx}
\begin{document}

\title{Short-range
coherence of a lattice Bose atom gas in the Mott insulating phase
}

\author{Yue Yu }
\affiliation{Institute of Theoretical Physics, Chinese Academy of
Sciences, P.O. Box 2735, Beijing 100080, China}
\date{\today}
\begin{abstract}
We study the short-range coherence of ultracold lattice Bose gases
in the Mott insulating phase. We calculate the visibility of the
interference pattern and the results agree quantitatively with the
recent experimental measurement (Phys. Rev. Lett. {\bf 95}, 050404
(2005)). The visibility deviation from the inversely linear
dependence on the bare on-site interaction $U_0$ is explained both
in smaller and larger $U_0$. For a smaller $U_0$, it comes from a
second order correction. For a larger $U_0$, except the breakdown
of adiabaticity as analyzed by Gerbier et al, there might be
another source to cause this deviation, which is the diversity
between $U_0$ determined by the single atom Wannier function and
the effective on site interaction $U_{\rm eff}$ for a
multi-occupation per site.
\end{abstract}


\maketitle

The observation of the Mott insulating phase in ultracold Bose
gases in an optical lattice opens a new era to investigate exactly
controllable strong-correlated systems \cite{jaks,mott}. For a
one-component lattice Bose gases, the Bose Hubbard model \cite{BH}
captures the basic physics of the systems \cite{jaks}. The
theoretical studies mostly focused on the sharp phase transition
between the superfluid/Mott-insulator
\cite{mean,oosten,num,dk,pv,yu,lu}. This phase transition may play
an important role in various quantum information processing
schemes \cite{quif}.

Recently, the residual short-range interference in the insulating
phase has been predicted by numerical studies \cite{pred}. This
phase coherence has been observed by a measurement of the
visibility of the interference pattern\cite{gerb}. It was found
that the visibility is inversely proportional to the on-site
interaction strength $U_0$ of the Bose Hubbard model in a wide
range. In explaining their data, Gerbier et al assumed a small
admixture of particle-hole pairs in the ground state of the Mott
insulating phase. They showed that the visibility of interference
pattern calculated by this ground state may well match the
experimental data in a wide intermediate range of $U_0$.

There were deviations from the inverse linear power law in both
small and large $U$ in the measurement of the visibility.  Gerbier
et al interpreted the large $U$ deviation is caused by a breakdown
of adiabaticity since the ramping time used in the experiment has
been close to the tunnelling time. For the deviation in a small
$U$, there was no explanation yet \cite{new}.

In this paper, we will analytically prove the inverse linear power
law of the visibility for intermediate $U$ in the zero
temperature. Here the words 'intermediate $U$' (as well as 'small
$U$', 'large $U$' in this work) mean the magnitude of $U-U_c$ is
intermediate (small or large), with $U_c$ the critical interaction
strength of the superfluid/Mott insulator transition. The result
is exactly the same as that obtained by Gerbier et al by assuming
a small admixture of the particle hole pair in the ground state
\cite{gerb}. We also show the deviation of the visibility from the
inverse linear power law in a small $U$ is caused by a second
order correction. For the large $U$, we show that, except the
explanation by the authors of the experimental work, owing to the
multi-occupation per site, the effective on-site interaction
$U_{\rm eff}$ which appears in the Bose Hubbard model
\cite{cpt,oo,li} is different from $U_0$ which was determined by
the single atom Wannier function and used to fit the data of the
experiment.

We consider a one-component Bose gas in a 3-dimensional optical
lattice described by a periodic potential $V_0({\vec r})$.
Although the real experimental system was confined by a trap
potential, we here only pay our attention to the homogeneous
system. Beginning with the expansion of the boson field operators
in a set of localized basis, i.e., $\psi (\vec r )=\sum_i a_i
w(\vec r-\vec r_ i)$ and keeping only the lowest vibrational
state, one can define an on-site free energy $ f=\bar nI+U\bar
n(\bar n-1)/2, $ where $\bar n$ is the average occupation per
site. The on-site energy $I$ and the bare on-site interaction $U$
are defined by $I=\int d\vec r w^*(\vec r
)[-\frac{\hbar^2}{2m}\nabla^2+V_0(\vec r)] w(\vec r)$,
\begin{eqnarray}
U=\frac{4{\pi}a_s\hbar^2}{m}\int d\vec r|w(\vec r)|^4.
\end{eqnarray}
This  on-site free energy contributes to the chemical potential by
$\mu=-\partial f/\partial \bar n$ and defines the effective
on-site interaction \cite{cpt,oo,li}
\begin{eqnarray}
U_{\rm eff}=\partial^2 f/\partial \bar n^2.
\end{eqnarray}
For the single occupation per site, $U_{\rm eff}=U=U_0$ and the
difference appears for $\bar n>1$. We will be back to this issue
later. The Bose Hubbard model for a homogeneous lattice gases is
defined by the following Hamiltonian
\begin{eqnarray}
H=-t\sum_{\langle ij\rangle} a^\dag_ia_j+\frac{U_{\rm eff}}2\sum_i
a^\dagger_ia_ia^\dagger_ia_i-\mu\sum_ia^\dagger_ia_i,
\end{eqnarray}
where $\langle ij\rangle$ denotes the sum over the nearest
neighbor sites and $\mu$ is the chemical potential. The tunnelling
amplitude is defined by
$$
t_{B,F;ij}=\int d\vec r w^*(\vec r+\vec
r_i)[-\frac{\hbar^2\nabla^2}{2m}+V_0(\vec r)]w(\vec r+\vec r_j),
$$
for  a pair of the nearest neighbor sites $(i,j)$.

Our main goal is to calculate the interference pattern
\begin{eqnarray}
S(\vec k)=\sum_{i,j} e^{i\vec k\cdot (\vec r_i-\vec r_j)}\langle
a^\dag_ia_j\rangle,
\end{eqnarray}
which is related to the density distribution of the expanding atom
clouds by $\rho(\vec r)=\frac{m}{\hbar t_{ex}}|\tilde w(\vec
k=m\vec r/\hbar t_{ex})|^2S(\vec k)$ with $m$ the atom mass and
$t_{ex}$ the time of the atom free expansion \cite{pred,pe}. Since
we are interested in the Mott insulating phase, we can calculate
$S(\vec k)$ by taking the tunnelling term as a perturbation. To do
this, we introduce a Hubbard-Stratonovich field in the partition
function \cite{dk}
\begin{eqnarray}
&&Z[J,J^*]=\int {\cal D}\Phi^*{\cal D}\Phi {\cal D}a^*{\cal D}a
\exp\biggl\{-S_0\nonumber\\&&+t\int_0^\beta d\tau \sum_{\langle
ij\rangle}a^*_ia_j
+ \int_0^\beta d\tau\sum_i(J^*_ia_i+J_i a^*_i)\nonumber\\
&&-t\int_0^\beta
d\tau(a^*_i-\Phi^*_i+J^*_i/t)(a_j-\Phi_j+J_j/t)\biggr\},
\end{eqnarray}
where $S_0$ is the $t$-independent part in the full action and $J$
and $J^*$ are currents introduced to calculate correlation
functions. Integrating away $a_i$ and $a^*_i$ and transferring
into the lattice wave vector and thermal frequency space, one has
\begin{eqnarray}
Z[J^*,J]&=&\int {\cal D}\Phi^*{\cal D}\Phi\exp\biggl\{\sum_{\vec
k,n}(-\Phi_{\vec k,n}^*{\cal G}^{-1}(\vec k,i\omega_n)\Phi_{\vec k,n}\nonumber\\
&+&J^*_{\vec k,n}\Phi_{\vec k,n}+J_{\vec k,n}\Phi^*_{\vec
k,n}+\frac{1}{\epsilon_k}J^*_{\vec k,n}J_{\vec k,n},
\end{eqnarray}
where $\epsilon_k=-2t\sum_{\nu=x,y,z}\cos k_\nu$. The correlation
function is calculated in a standard way:
\begin{eqnarray}
&&\langle a^*_{\vec k,n} a_{\vec
k,n}\rangle=\frac{1}{Z[0,0]}\frac{\delta^2 Z[J^*,J]}{\delta
J^*_{\vec k,n}\delta J_{\vec k,n}}\biggr|_{J^*=J=0}\nonumber
\\
&&=\langle \Phi^*_{\vec k,n} \Phi_{\vec
k,n}\rangle+\frac{1}{\epsilon_k}=-{\cal G}(\vec
k,i\omega_n)+\frac{1}{\epsilon_k}.
\end{eqnarray}
The interference pattern then may be expressed as
\begin{eqnarray}
S(\vec k)=-\frac{1}\beta\sum_n[{\cal G}(\vec
k,i\omega_n)-\frac{1}{\epsilon_k}].
\end{eqnarray}
In the Mott insulating phase, the correlation function ${\cal
G}(\vec k,i\omega_n)$ has been calculated by slave particle
techniques \cite{dk,yu}
\begin{eqnarray}
{\cal G}^{-1}({\bf k},i\omega_n)=\epsilon_{\bf k}-\epsilon_{\bf
k}^2\sum_{\alpha=0}^\infty(\alpha+1)\frac{n^\alpha-n^{\alpha+1}}{i\omega_n+\mu
-\alpha U}, \label{green}
\end{eqnarray}
where the slave particle occupation number is given by
\begin{eqnarray}
n^\alpha=\frac{1}{\exp\{\beta[-i\lambda-\alpha\mu
+\alpha(\alpha-1)U_{\rm eff}/2]\}\pm 1}, \label{oc}
\end{eqnarray}
which obeys $\sum_\alpha n^\alpha=1$ and $\sum_\alpha \alpha
n^\alpha=N$ in the mean field approximation \cite{note}. $\lambda
$ is a Lagrangian multiplier to ensure $\sum_\alpha n^\alpha=1$.
The sign $\pm$ corresponds to the slave fermion or boson,
respectively. In previous works, we have show that the slave
fermion approach may have some advantages to the slave boson
approach\cite{yu,lu}. We then take the slave fermion formalism. In
the Mott insulating phase, since $U_{\rm eff},\mu\gg t$, one can
expand ${\cal G}(\vec k,i\omega_n)$ in terms of
$\epsilon_k/(i\omega_n-\mu+\alpha U_{\rm eff})$ and the
interference pattern reads
\begin{eqnarray}
S(\vec k)&=&-\frac{1}\beta\sum_n({\cal G}(\vec k,\omega_n)-\frac{1}{\epsilon_k})\nonumber\\
&=&
\frac{1}\beta\sum_n\sum_{a=0}(-1)^a\epsilon_k^a(A(\omega_n))^{a+1}
,\label{skf}\\
A(\omega_n)&=&\sum_{\alpha=0}^\infty(\alpha+1)\frac{n^{\alpha+1}
-n^\alpha}{i\omega_n+\mu-\alpha U_{\rm eff}}.\nonumber
\end{eqnarray}

\begin{figure}[htb]
\begin{center}
\includegraphics[width=8cm]{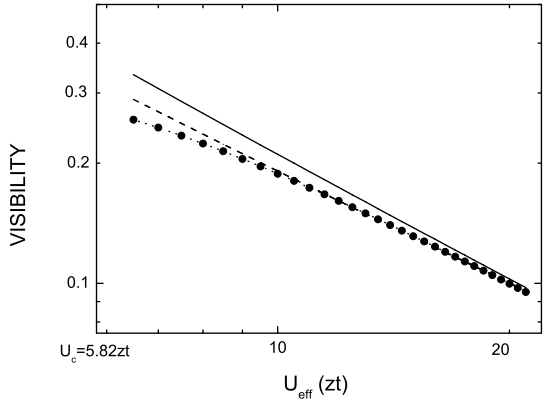}
\end{center}
\vspace{-4mm}
 \caption{Visibility of the interference pattern versus
 $\bar U_{\rm eff}$ according to (\ref{vi}) in a log-log plot(the dot line with circles)
 .
 The solid line is the inverse linear power law (\ref{lin}) and
 the dash line is a power law fit with an exponent $-0.95$
 to (\ref{vi}). }
 \label{1}
\end{figure}

Making the frequency sum, one has, to the first order of
$\epsilon_k$,
\begin{eqnarray}
S(\vec k)&\approx&-\sum_\alpha n_B(\alpha U_{\rm
eff}-\mu)(\alpha+1)(n^{\alpha+1}-n^\alpha)
\nonumber\\
&-&\epsilon_k\beta\sum_\alpha[(\alpha+1)^2(n^{\alpha+1}-n^\alpha)^2
\nonumber\\
&\times&n_B(\alpha U_{\rm eff}-\mu)(1+n_B(\alpha U_{\rm eff}-\mu))
\label{sk}
\\
&-&\frac{2\epsilon_k}{U_{\rm eff}}\sum_{\alpha<\gamma} (n_B(\alpha
U_{\rm eff}-\mu)-n_B(\gamma
U_{\rm eff}-\mu))\nonumber\\
&\times&(\alpha+1)(\gamma+1)(n^{\alpha+1}-n^\alpha)(n^{\gamma+1}-n^\gamma)
,\nonumber
\end{eqnarray}
where $n_B(\alpha U_{\rm eff}-\mu)=[e^{\beta(\alpha U_{\rm
eff}-\mu)}-1]^{-1}$. In the limit $T\to 0$ and the $n_0$-th Mott
lobe, one knows $(n_0-1)U_{\rm eff}<\mu<n_0 U_{\rm eff}$ and
$n^\alpha=\delta_{\alpha,n_0}$. Substituting these into
(\ref{sk}), one obtains the zero temperature value of $S(\vec k)$
\begin{eqnarray}
S(\vec k,T=0)=n_0-2n_0(n_0+1)\frac{\epsilon_k}U_{\rm eff}.
\end{eqnarray}
This is what Gerbier et al obtained by assuming the particle-hole
pair admixture in the ground state \cite{gerb}. Integrating along
one lattice direction, the corresponding 2D visibility is given by
\begin{eqnarray}
{\cal V}=\frac{\rho_{\rm max}-\rho_{\rm min}}{\rho_{\rm
max}+\rho_{\rm min}}= \frac{S_{\rm max}-S_{\rm min}}{S_{\rm
max}+S_{\rm min}}\approx \frac{4}3(n_0+1)\frac{zt}U_{\rm eff}
\label{lin}
\end{eqnarray}
 for $z=6$, where
$\rho_{\rm max}$ and $\rho_{\rm min}$ are chosen such that the
Wannier envelop was cancelled. This is the inverse linear power
law used to fit the experimental data \cite{gerb}. However, the
experimental data deviated from this power law fit when $U_{\rm
eff}/zt<8$. In terms of (\ref{skf}), we think that this comes from
a second order correction. A direct calculation shows that the
second order correction in zero temperature is given by
\cite{note1}
\begin{eqnarray}
 \delta^{(2)} S(\vec
k)=3n_0(n_0+1)^2\frac{\epsilon_k^2}{U_{\rm eff}^2}.\label{sk0}
\end{eqnarray}
 Thus, the 2D
visibility for $n_0=1$ is modified to
\begin{eqnarray}
{\cal V}=\frac{8}{3\bar U_{\rm eff}(1+32\bar U_{\rm
eff}^{-2}/3)},\label{vi}
\end{eqnarray}
with $\bar U_{\rm eff}=U_{\rm eff}/zt$. In Fig. \ref{1}, we show
the visibility against $\bar U_{\rm eff}$ in a log-log plot for
$n_0=1$. This second order correction suppresses the visibility
for a small $\bar U_{\rm eff}$ while the exponent of the power law
seems deviating from $-1$ a little. These features agree with the
experimentally measured data.

We have neglected the finite temperature effect to compare with
the experiment although our theory is in finite temperature. In
fact, there may be a finite temperature correction to the
interference pattern in the second order. According to
(\ref{skf}), it is given by, near $n_0=1$
\begin{eqnarray}
\delta ^{(2)}S_T(\vec k)=18(n^1)^2n^2\frac{\epsilon^2_k}{ U_{\rm
eff}^2},\label{skt}
\end{eqnarray}
which may further suppresses the visibility. For instance, at
$T=1.0zt\sim 10^1$nK, the ratio between (\ref{skt}) and
(\ref{sk0}) is
\begin{eqnarray}
&&\frac{\delta ^{(2)}_TS(\vec k)}{\delta ^{(2)}S(\vec
k)}=3(n^1)^2n^2/2\nonumber\\&&=0.106,0.098,0.087,~{\rm and}~ 0.064
\nonumber
\end{eqnarray}
 for $\bar
U_{\rm eff}=6,7,8,$ and 10. However, the temperature in the Mott
insulator is difficult to be estimated in the experiment
\cite{thank}. Thus, a quantitative comparison of the finite
temperature calculation to the experiment data is waiting for more
experimental developments.

We now discuss the large $U$ deviation from the inverse linear
power law, which has been seen in the experiment and explained by
the breakdown of adiabaticity \cite{gerb}. We will reveal another
possible source for this deviation. As we have mentioned before,
the value of $U_{\rm eff}$ may be different from $U$ and $U_0$ for
$\bar n>1$. Our above calculation showed an inverse linear power
law to $U_{\rm eff}$ whereas the experimentalists used $U_0$ to
fit their data.

\begin{figure}
\vspace{-3mm}
\begin{center}
\includegraphics[width=7cm]{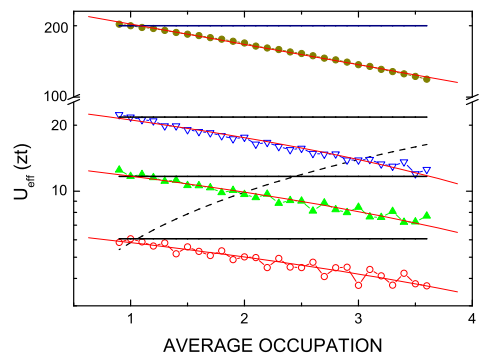}
\end{center}

\vspace{-7mm}

 \caption{(Color online) The effective on-site interaction $U_{\rm eff}$ versus
 the average occupation per site, $\bar n$ in a $\bar n$-$\log(U_{\rm eff})$ plot. The thin solid lines
 are linear fits to variational data for $V_0=11.95,14.32$,
 16.25 and 29 $E_R$ (empty circles, filled triangles, empty
 triangles
 and filled circles, respectively). The dash line is critical
 interaction strength calculated by the mean field theory
 \cite{mean,oosten}. The thick horizontal lines are the on-site
 interactions $U_0$ calculated by the single atom Warrier function.
 }\label{2}
\end{figure}

Due to the interaction, the atom energy band may be modified and
the Wannier function may be broadened, compared to the single atom
ones. In Ref. \cite{li}, we have considered the mean field
interaction and made a variational calculation to the Wannier
function by using Kohn's method \cite{kohn}. The direct result of
the broadening of the Wannier function is the bare on-site
interaction $U$ becomes weaker than $U_0$ which is calculated by
the single atom Wannier function. The $\bar n$-dependence of $I$
may further reduce $U_{\rm eff}$ from $U$. In Fig. \ref{2}, we
plot $U_{\rm eff}$ versus $\bar n$. In the low part of Fig.
\ref{2}, three typical lattice depths
 are considered, $V_0=11.95,14.32$ and
16.25 $E_R(=\frac{\hbar^2 k^2}{2m})$, corresponding to the
critical interaction strengths of the $n_0=1,2$ and 3 Mott states.
The up-part is for $V_0=29E_R$, which was the lattice depth where
the adiabaticity breaks \cite{gerb}.

Several points may be seen from Fig. \ref{2}. First, the critical
values of $V_0=14.32 E_R$ for $n_0=2$ and $16.25 E_R$ for $n_0=3$
are closer to experimental ones, 14.1(8) $E_R$ and 16.6(9) $E_R$
\cite{gerb}, comparing to 14.7 $E_R$ and 15.9$E_R$, corresponding
to the single atom Wannier functions. Second, the variational data
are downward as $\bar n$ indicates that $-\log U_{\rm eff}>-\log
U_0$ for $\bar n>1$. This may cause two results: (a) If $-\log
U_{\rm eff}$ deviates from $-\log U_0$ a small magnitude, the
power law fit presents an exponents $-(1-\delta)$. This has been
observed in experiment, which is $-0.98(7)$ \cite{gerb}. (b) As
$\bar n$ increases, the deviation becomes significant. This may
appear in a large $V_0$. In the experiment, the latter appeared in
$V_0>29 E_R$. We show that, in Fig. \ref{2}, the deviation is not
a small magnitude for $V_0=29 E_R$.

In summary, we studied the short-range coherence in the Mott
insulating phase with a finite on-site interaction strength. The
interference pattern and then its visibility were calculated by
using a perturbation theory. The inverse linear power law of the
visibility to the interaction strength, which was found in the
experiment, was exactly recovered. We further discussed the
deviation from this power law both in a small and large $U_0$. We
found that a second order effect suppresses the visibility for a
small $U_0$ while its up-deviation in a large $U_0$ might be
caused by the difference between $U_0$ and $U_{\rm eff}$ except
the possible breakdown of adiabaticity.

 We would like to thank Fabrice Gerbier for useful discussions.
 This work was supported in part by the National Natural
 Science Foundation of China.


\begin{references}
\bibitem{note} We do not consider the fluctuations to these two
equations.
\bibitem{note1} There is also a correction $\delta^{(2)} S\propto \bar U_{\rm eff}^{-2}$ for $n_0>1$ in principle. However,
this corresponds to a much larger $\bar U_{\rm eff}$ and
therefore, may become not visible in experiments.
\end{references}
\end{document}